\documentclass[prl,twocolumn, showpacs]{revtex4}
\usepackage{graphicx}
\usepackage{color}

\begin{document}
\title{Enhancing the Detection of Natural Thermal Entanglement with Disorder}
\author{Jenny Hide$^1$, Wonmin Son$^{1,2}$ and Vlatko Vedral$^{1,2,3}$}
\affiliation{Department of Physics and Astronomy, EC Stoner Building, University of Leeds, Leeds, LS2 9JT, UK$^1$ \\
Centre for Quantum Technologies, National University of Singapore, 3 Science Drive 2, Singapore 117543$^2$ \\
Department of Physics, National University of Singapore, 2 Science Drive 3, Singapore 117542$^3$.}

\date{\today}

\begin{abstract}
Physical systems have some degree of disorder present in them.
We discuss how to treat natural, thermal entanglement in any
random macroscopic system from which a thermodynamic
witness bounded by a constant can be found. We propose that
functional many-body perturbation theory be applied to allow
either a quenched or an annealed average over the disorder to be
taken. We find when considering the example of an $XX$ Heisenberg
spin chain with a random coupling strength, that the region of
natural entanglement detected by both witnesses can be enhanced by
the disorder.

\end{abstract}

\maketitle



Many-body systems have been of great interest to the condensed matter
community for numerous years. More recently, ideas from quantum
information have been applied to such systems, allowing new methods of
quantifying their properties to be utilised. Entanglement in many-body systems,
\cite{manybody} is of particular significance due both to fundamental
interest and applications in quantum computation. Approaching the
thermodynamic limit and at finite temperatures, this \emph{natural}
entanglement becomes relevant in real, macroscopic systems \cite{macroent}.

Although entanglement between any two qubits, or even subsystems,
in a large system can be (comparatively)
easily calculated, quantifying entanglement between more qubits
is much more complex \cite{manybody}. However, using an entanglement witness
we can still find critical values of the system's parameters
below which the system is entangled. An entanglement witness is an
observable whose expectation value is bounded
for any separable state. Using a witness is advantageous as we can
detect natural, and therefore thermal,
entanglement of all classes while other measures detect only bipartite
entanglement and/or can be used only at zero temperature.

When an entanglement witness can be derived from thermodynamic
quantities, in particular the partition function, we refer to
it as a thermodynamic witness \cite{dowling}. If a system contains
some randomness, an average over this disorder must be performed.
The method used to calculate the average
is dependent on the time taken to measure the witness in an experimental
situation, giving either a \emph{quenched} or an \emph{annealed}
average \cite{quench}. For each type of average, the witness will detect a
different region of entanglement. It is therefore important
to distinguish which we are able to measure experimentally.
Previous research into average entanglement \cite{Chiara,random1,binosi}
has concentrated on spin chains and used sampling, which corresponds to
calculating
a quenched average numerically, or a renormalisation group approach to
calculate how the randomness in the chain affects entanglement. These
methods cannot take the differences between quenched and
annealed entanglement into account. In addition, these papers
look at zero temperature, bipartite entanglement. It is both in these
areas, and by introducing a perturbative technique to deal with disorder
in the context of entanglement, that our work offers new insight.

This letter studies natural entanglement in random systems. After
averaging over the disorder, the regions of
entanglement detected by the quenched and annealed witnesses can be
related by Jensen's inequality \cite{jensen}.
As these averages are difficult to perform on a large system,
we introduce the concept of using the well established
many-body perturbation theory to calculate the thermodynamic entanglement
witness. This method is very general and can be used, at least
numerically, for any system.

To exemplify how these different averages affect a system's
entanglement, we consider an $XX$ spin half chain in a
uniform magnetic field with an extra random coupling strength.
This system in particular
is used as the majority of the calculation can be done
analytically rather than numerically. Hence we calculate both the
quenched and annealed witnesses for this system using many-body
perturbation theory and plot the results. We find that the randomness
can both enhance and create regions of entanglement detected by both
witnesses.
We stress that it is the region of entanglement detected by
the witness that
is enhanced by the disorder, rather than, necessarily, the entanglement
itself. Intuitively however, increasing the disorder can increase the
entanglement itself, hence we discuss possible mechanisms for this.


In order to calculate an entanglement witness in a random system,
we must perform an average over the disorder.
As discussed above, the appropriate average to
use is dependent on how fast the measurement process is in comparison
to the speed of the fluctuations of the variable in the system.
As different experimental conditions require different averages to
be taken of the system, it is important to determine which is used.

First we consider the possibility that we can measure each state
of the system faster than the disorder fluctuates.  This is
called a \emph{quenched} average. To calculate this, we first
perform a thermal average, and then average over the random
variable. Rather than calculating the properties of the system,
we calculate an average of all possible systems.
Otherwise, if the system fluctuates faster than we can measure it,
we cannot measure each state the system takes, and we should
use an \emph{annealed} average. Here, the system thermalises
before we can measure the state; hence the average over the
disorder and the thermal average are taken at the same time.
Experimental studies on high temperature superconducting materials
discuss situations when each of these averages occur \cite{experi}.

As each type of average relates to a different standard
of measurement, each will give a different region of entanglement.
It is therefore important to clarify which is needed when measuring a
system. The regions of entanglement detected by the quenched and
annealed averages can be related by invoking Jensen's inequality,
$\langle f\left( X \right) \rangle \leq f \langle X \rangle$,
where is $f$ is a concave function acting on $X$ and the
average, $\langle \cdots \rangle$ is over the disorder. For a convex
function, the inequality is reversed. The left hand side of the inequality
is the quenched average, and the right hand side is the annealed average.
We are interested in how these averages can change the amount of
entanglement we can measure, hence we must consider how we can use
entanglement witnesses in this context.

For any system of finite dimension for which we can calculate a
thermodynamic witness with some constant value, $\alpha$, as
the bound for detecting entanglement, we can apply the following
argument. If the witness, $W$, is a convex function, the entangled region
detected which we define as $E=\alpha -W$ is always concave. Similarly,
$E$ is convex if the witness is concave. Hence the application of
Jensen's inequality to either scenario tells us which average
allows us to detect the most entanglement.

Although entanglement has previously been studied using
thermodynamic witnesses calculated from the partition function of the
system, to discover how disorder affects thermal, macroscopic (natural)
entanglement, we instead use functional many-body
perturbation theory.

We set out an outline of the method for a fermionic system below. Due to
the Jordan-Wigner transformation, this also applies to any spin half system.
Bosonic systems can also be studied using this formalism, but
complex numbers are used in place of Grassmann
variables. We first define a generating functional, $Z[\bar{\eta},\eta]
|_{\bar{\eta},\eta=0}= Z$ where $Z=Tr(e^{-\beta H})$ is the
partition function of the
system and $\eta$ is a Grassmann variable. $Z[\bar{\eta},\eta]$ allows
us to calculate expectation values using
functional derivatives from an appropriate generating function, $g_f$:

\begin{equation}
Z\left[\bar{\eta},\eta \right]= Tr\left(e^{-\beta H} T e^{g_f[\bar{\eta},
\eta]}\right)
\label{eq:Z_eta}
\end{equation}

\noindent where $\beta$ is the inverse temperature,
$H=H_0+H_1(\upsilon_l)$
is the Hamiltonian of the system and $\upsilon_l$ is a random variable,
with $l$ labelling position. $T$ is the time ordering operator.
After putting (\ref{eq:Z_eta}) into path integral
representation \cite{perturb}, we can take an average over the
random variable of $Z[\bar{\eta},\eta]^n$. For
clarity, we consider the random variables to be taken from a
Gaussian distribution.

\begin{equation}
\left\langle Z^n\left[\bar{\eta},\eta \right] \right\rangle
=\frac{1}{(2 \pi \Delta)^{\frac{N}{2}}} \int_{-\infty}^\infty
\prod_{l=1}^N d\upsilon_l e^{-\frac{1}{2\Delta}\sum_l \upsilon_l^2}
\prod_{a=1}^n
Z\left[\bar{\eta}_a,\eta_a \right]
\label{eq:ave_Z1}
\end{equation}

\noindent where $\Delta$ is the variance. After calculating this average, we
perform transformations to diagonalise $H_0$. The average can then
be rewritten as

\begin{equation}
\left\langle Z^n\left[\bar{\eta},\eta \right] \right\rangle =
Z_0^n e^{\Delta \tilde{S}^n_{int}[\frac{\delta}{\delta \bar{\eta}},
\frac{\delta}{\delta \eta}]} e^{g_F[\bar{\eta},\eta]}
\label{eq:ave_Z2}
\end{equation}

\noindent where $Z_0$ is the partition function of the unperturbed system,
$\Delta \tilde{S}^n_{int}[\frac{\delta}{\delta \bar{\eta}},
\frac{\delta}{\delta \eta}]$ is the perturbed term, and acts on
$g_F[\bar{\eta},\eta]$, the generator re-written in terms of a
Greens function. Expanding $e^{\Delta \tilde{S}^n_{int}}$, we
can use Feynman diagrams to calculate $\langle Z^n[\bar{\eta},
\eta]\rangle$. Functional derivatives can then be used to calculate
an entanglement witness for both the quenched and the annealed
averages. We can then implement the
replica trick, $\langle \ln Z \rangle = \lim_{n\rightarrow 0}\frac{1}{n}
(\langle Z^n \rangle - 1)$ in order to calculate the quenched witness.
The annealed average is found simply by setting $n=1$. The replica trick
corresponds to considering $n$ identical, non-interacting replicas of the
system. Performing an average over the disorder of these replicas allows
us to use $\Delta$, the variance rather than a different
random variable at each $l$. Rather than considering the experimental
situation of measuring the system many times and performing an average,
the replica trick allows us to instead use many copies of the same system.

Perturbation theory is valid when the matrix elements,
$\langle m | H_1 (\upsilon_l) | n \rangle$, of the interaction term,
$H_1 (\upsilon_l)$, where $|n\rangle$ and $|m\rangle$ ($m\neq n$) are
possible eigenvectors of $H_0$, are much smaller than the energy
difference between $n$ and $m$. To ensure this condition is satisfied,
we require $\upsilon_l \ll 1$ with high probability, and hence $\Delta \ll 1$.



To make the arguments above explicit, we calculate the average witness
for a specific, predominantly analytically solvable, example: an $XX$
spin half chain with a random coupling strength. The unperturbed
Hamiltonian is

\begin{equation}
H_0= -\sum_{l=1}^N \left[ \frac{J}{2}\left( \sigma_l^x \sigma_{l+1}^x +
\sigma_l^y \sigma_{l+1}^y \right) + B\sigma_l^z \right]
\label{eq:H_0}
\end{equation}

\noindent where $J$ is the coupling strength between nearest
neighbours, and $B$ is a uniform magnetic field. Considering the
thermodynamic limit, $N\rightarrow
\infty$, the spin chain can be diagonalised \cite{katsura}
using a Jordan-Wigner transformation $a_l=\prod_{m=1}^{l-1} \sigma_m^z
\otimes \left(\sigma_l^x+i\sigma_l^y \right)/2$ and a Fourier transform,
$a_l^{} =\int_{-\pi}^\pi \frac{dq}{2\pi} e^{iql} d(q)$
giving $H_0=\int_{-\pi}^\pi \frac{dq}{\pi} (J\cos q-B) d^\dagger (q)
d(q)$. The partition function is then

\begin{equation}
\ln Z_0 = \frac{N}{2\pi} \int_{-\pi}^\pi dq
\hspace{1mm} \ln \left(2\cosh\left[ \beta\left( J\cos q -B\right)
\right] \right)
\label{eq:partfn}
\end{equation}

\noindent We now introduce the Hamiltonian $H=H_0+H_1(\upsilon_l)$ and
consider $H_1(j_l)= -\frac{1}{2}\sum_{l=1}^{N-1} j_l\left( \sigma_l^x
\sigma_{l+1}^x + \sigma_l^y \sigma_{l+1}^y \right)$ where $j_l$ is
a site dependent random coupling strength between nearest neighbours.
Each random variable is taken from a Gaussian distribution
centred at zero with variance $\Delta$.

To detect entanglement, we use an entanglement witness \cite{hide}
which is proportional to the sum of the $x$ and $y$ components of
the susceptibility, and is therefore measurable in practice \cite{susc}.

\begin{equation}
W=\left| \frac{2}{\beta N} \frac{\partial}{\partial J} \ln Z \right|
=\left|\frac{1}{N} \sum_l \langle \sigma_l^x \sigma_{l+1}^x + \sigma_l^y
\sigma_{l+1}^y \rangle \right|
\label{eq:witness}
\end{equation}

\begin{figure}[t]
\begin{center}
\centerline{
\includegraphics[width=3.0in]{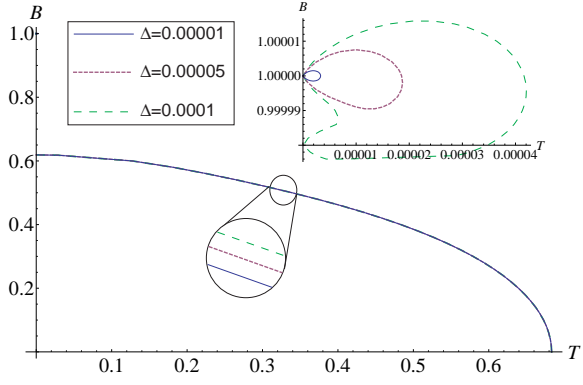}
 }
\end{center}
\caption{This plot shows the values of uniform magnetic field, $B$
and temperature, $T$, with $J=1$ below which entanglement is detected
by the quenched witness for various values of the disorder, $\Delta$.
Details of this plot are discussed in the text.
} \label{fig1}
\end{figure}

\noindent The system is entangled when the witness is greater than 1.
This bound is found using the Cauchy-Schwarz inequality and properties
of the density matrix \cite{toth,ved1}.

The entangled region, $E=\alpha-W$, is concave as $W$ is always convex
due to the absolute sign.
Using Jensen's inequality, this means the region of entanglement detected
by the quenched witness is equal to or less than the region detected
by the annealed witness.


In order to use functional many-body perturbation theory, we must first
define an
appropriate generating function with which to calculate the witness.
Letting $g_f[\bar{\eta},\eta]=\int_0^\beta d\tau \sum_l \left[
\bar{\eta}_l(\tau) a_l(\tau) + \eta_l(\tau) a_l^\dagger(\tau)\right]$
in (\ref{eq:Z_eta}) and diagonalising the witness,

\begin{equation}
W  = \left| \frac{4}{N}\int_{-\pi}^\pi \frac{dq}{2\pi} \cos q \langle
d^\dagger (q) d (q) \rangle \right|
\label{eq:witness_diag}
\end{equation}

\noindent we find $\langle d^\dagger (q) d (q) \rangle= \lim_{n
\rightarrow x} \frac{\delta}{\delta \eta_b(\tau',q')}
\frac{\delta}{\delta \bar{\eta}_a(\tau,q)} \langle Z^n
[\bar{\eta},\eta] \rangle$ with $x=0$ for the quenched average and
$x=1$ for the annealed average.

Considering $H_1(j_l)$, we average over $j_l$ in (\ref{eq:ave_Z1})
and calculate equation (\ref{eq:ave_Z2}) from $\ln Z_0$ as
given in (\ref{eq:partfn}), $\Delta \tilde{S}^n_{int} =
\frac{1}{2}\int_0^\beta d\tau d\tau' \int_{-\pi}^\pi \Pi_{j=1}^4
\frac{dq_j}{2\pi} \sum_{abcd} \delta_{ab} \delta_{cd} 2\pi
\delta(q_1-q_2+q_3-q_4)
\left( e^{iq_1}+e^{-iq_2} \right)\left( e^{iq_3}+e^{-iq_4}\right)
\left[ 2\pi \frac{\delta}{\delta \bar{\eta}_b(\tau,q_2)} 2\pi
\frac{\delta}{\delta \eta_a(\tau,q_1)} \right].$ $\left[ 2\pi
\frac{\delta}{\delta \bar{\eta}_d(\tau',q_4)} 2\pi
\frac{\delta}{\delta \eta_c(\tau',q_3)} \right]$,
and the source term,
$g_F[\bar{\eta},\eta]=\int_0^\beta d\tau d\tau'
\int_{-\pi}^\pi \frac{dq}{2\pi} \sum_{ab} \bar{\eta}_a(\tau,q) G_{ab}(\tau
-\tau',q) \eta_b(\tau',q)$.

After expanding $e^{\Delta \tilde{S}^n_{int}}$ and calculating
Feynman diagrams to first order, we find the witness is

\begin{equation}
W_{ave}=\left| \frac{2}{\pi}\int_{-\pi}^{\pi} dq \cos q \left[n(q) -
\Delta G_{ave}^{(1)} \right] \right|
\end{equation}

\noindent with $ave$ indicating either a quenched, $q$ or annealed,
$a$ average.
From the method outlined above, we find that $G_q^{(1)}=\frac{2}{\pi}
\int_{-\pi}^{\pi} dp \cos^2 \left(
\frac{q+p}{2}\right)\left[ \frac{n(q)-n(p)}{\left(\varepsilon(p)-
\varepsilon(q) \right)^2} - \frac{\beta n(q) \left(1-n(q) \right)}
{\varepsilon(p)-\varepsilon(q)} \right]$, and
$G_a^{(1)}=G_q^{(1)}+\frac{2\beta^2}{\pi}n(q)(1-n(q))
\cos q\int_{-\pi}^{\pi} dp \hspace{1mm} n(p)\cos p$. Here,
$n(x)=\left( e^{\beta \varepsilon(x)}+1\right)^{-1}$
and $\varepsilon(x)=2J\cos x-2B$.

\begin{figure}[t]
\begin{center}
\centerline{
\includegraphics[width=3.0in]{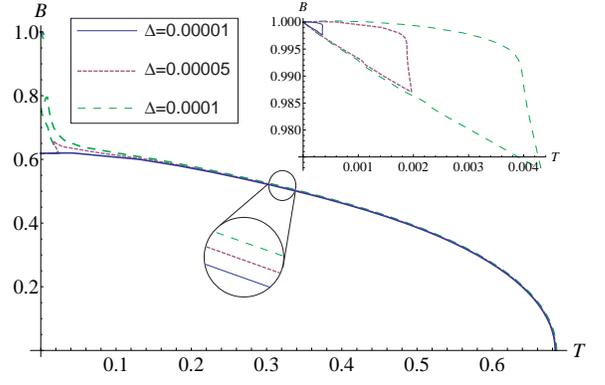} }
\end{center}
\caption{This plot shows the values of uniform magnetic field, $B$
and temperature, $T$, with $J=1$ below which entanglement is detected
by the annealed witness for various values of the disorder, $\Delta$.
Details of this plot are discussed in the text.}\label{fig2}
\end{figure}

As discussed previously, to ensure our use of perturbation theory is valid,
we restrict the value of the disorder to $\Delta \leq 0.0001$.
We note that all the features found in Figs. (\ref{fig1}) and
(\ref{fig2}) are present even when $\Delta \ll 0.0001$.

The entanglement detected by the witness in an unperturbed $XX$ spin half
chain, plotted in \cite{hide} (Fig. (1) at $b=0$), exists
below a critical uniform magnetic field, $B$, and temperature, $T$,
and is below the coupling strength between nearest neighbour sites,
$J$. Much attention has been focused on the study of quantum phase
transitions \cite{osterloh} and we note that $H_0$ has one at $B=J$.
The spectrum of the unperturbed spin chain \cite{son} at
zero temperature is as follows: at $B=0$, the ground state of the
spin chain is a fully symmetric state of all permutations of an equal
number of up and down spins. On increasing $B$, the
number of up spins increases until, when $B>J$, they are all spin up
and the system is in a product state.

Fig. (\ref{fig1}) shows how the entanglement detected by the
quenched witness behaves. Introducing the random coupling strength
enhances the region detected by the witness. Increasing $\Delta$
increases the entangled region detected compared to that
in the original $XX$ model. A
higher variance increases the probability of a high value of
$j_l$ (though this may be positive or negative) being picked
out of the Gaussian distribution. As we consider small $\Delta$
and a larger coupling strength increases entanglement, so does
increasing $\Delta$. Further, a region of
entanglement close to $B=J$ is created at low temperatures
which again increases with $\Delta$. We note that this extra region
of entanglement is created at the quantum phase transition of
$H_0$. The annealed witness is plotted in Fig.
(\ref{fig2}) and gives a larger entangled region than the quenched
witness everywhere except at $T=0$ where they coincide. Hence the plots
demonstrate the importance of an accurate determination of how we
measure the system experimentally. The behaviour of the plot as
$\Delta$ changes remains the same as
for the quenched witness.

Alternatively, we can view the system in the following way:
even if the ground state of a system is separable, there is a
possibility of creating entanglement by increasing the
temperature \cite{vlatko}. A finite temperature allows higher
energy levels to mix with the ground state. If these newly available
energy levels are themselves entangled, on mixing with a
separable ground state, entanglement can be created and/or increase.
Similarly, introducing disorder into a system allows higher energy
levels to be populated even at zero temperature, allowing for the
possibility of creating or enhancing entanglement.

We have also considered the effect of a random magnetic field,
$b_l$ on the witness.
The plots for the quenched and the annealed witness have similar features
as for $j_l$. The region of entanglement close to $B=J$ increases with
$\Delta$ similarly to the random coupling strength plots. At lower $B$,
increasing $\Delta$ decreases the entangled region, contrary
to the random coupling strength case. A random magnetic field increases
the chance of spin flip events. For lower $B$ this can only act to
decrease the entanglement
due to the highly entangled structure of the ground state of $H_0$.
At higher $B$, particularly at $B=J$, this same
effect will increase the entanglement due to the almost separable
ground state at this point.

This letter has studied entanglement in the limit of small disorder.
The opposite, large disorder limit is also of interest; for $b_l$,
it is clear that large disorder would cause the system to become separable
as $H\rightarrow \sum b_l \sigma_l^z$. However, the case for $j_l$ is
much more complex. Using the path integral formalism, we can consider
$H_1(j_l)$ to be the unperturbed part of the Hamiltonian, and $H_0$
to be the perturbation. As this cannot be done analytically, we
do not pursue this here. We do, however, still expect that in the
limit of large disorder the system should
become completely mixed, and hence separable, though the effect of averaging
over this disorder on the behaviour of the witness is nontrivial.
This uncertainty is due to the equally high probability of a large
positive, and hence
entanglement enhancing, or negative, and hence entanglement destroying,
value of $j_l$. While beyond the scope of this letter, this
would be an interesting topic to study.


We have discussed how the quenched and annealed entanglement witnesses
are related both experimentally and via Jensen's inequality, and found
that we can apply
functional many-body perturbation theory to a random system in
order to calculate these witnesses. We have found that randomness
in a system can enhance the entanglement detected by a witness.
In particular, we have considered a spin chain with a random
coupling strength. This result is important in the experimental realisation
of quantum systems due to entanglement being a resource in
quantum computation as it demonstrates that disorder may actually be
beneficial even in thermal systems.

\emph{Acknowledgements} - We are very grateful to I. Lawrie for many
extensive discussions on issues related to this letter. V.V.
and J.H. acknowledge the EPSRC for
financial support. V.V. is grateful for funding from the Wolfson
Foundation, the Royal Society and the E.U. His work is
also supported by the National Research Foundation
(Singapore) and the Ministry of Education (Singapore).


\end{document}